\documentclass[english,keywords,amsmath,amssymb,onecolumn]{revtex4}
\usepackage[T1]{fontenc}
\usepackage[latin1]{inputenc}
\usepackage{braket}
\usepackage{babel}%
\usepackage{graphicx}
\usepackage{color}
\usepackage{bm}
\usepackage{longtable}
\usepackage{amsmath}
\usepackage{amsfonts}
\usepackage{dsfont}
\usepackage{amssymb}
\usepackage{hyperref}
\begin{document}
\title{The 1964 paper of John Bell}
\author{Ujjwal Sen}
\affiliation{Harish-Chandra Research Institute, A CI of Homi Bhabha National Institute, Chhatnag Road, Jhunsi, Allahabad 211 019, India}

\begin{abstract}
We present a commentary on the famous 1964 paper of John Bell that rules out the entire class of underlying hidden variable theories  for quantum mechanics that are local.\end{abstract}

\maketitle
\section{Introduction}

Let us state at the onset that it is far better to directly read the 1964 article of John Stewart Bell~\cite{soja-hiseb} instead of the commentary  presented below. Let us also point the reader to the beautiful review of Mermin on this subject~\cite{Jaisalmer}. For a more updated and extensive review, see the one by Brunner, Cavalcanti, Pironio, Scarani, and Wehner~\cite{gojikoT}.
See also the recent essay by {\.Z}ukowski~\cite{dnaRa-ekTa-pulish-Daki}.

In 1935, Albert Einstein, Boris Podolsky, and Nathan Rosen had put forward an argument~\cite{aro-soja-hiseb} that quantum mechanics is ``incomplete''. They incorporated certain premises of ``locality'' and ``reality'' for any physical theory, 
and showed that quantum mechanics is incompatible with the premises. The paper of John Bell shows that the conjunction of the assumptions of locality and reality are incompatible with the statistical predictions of quantum mechanics. 

Einstein, Podolsky, and Rosen had used the language of continuous variables for describing their thought experiment, using Dirac delta ``functions'' among other tools. This was reformulated by Bohm and Aharonov~\cite{Parsvanath-mandir} using the algebra of quantum spins. Bell states and proves his result in this language of finite-dimensional spaces. 

The present contribution is confined to a brief commentary of the 1964 paper of John Bell. Please look at the contribution by N.D. Hari Dass in this issue for knowing about several other research topics into which John Bell delved into, as well as a brief biographical sketch of the scientist.

\section{The premises and the proof}

Let us begin by defining the premises of reality and locality.\\ 

\noindent \textbf{The reality assumption or the assumption of hidden variables} presupposes that measurements merely uncover the results that they provide. I.e., the observable that are being measured is assumed to already possess the value uncovered, in the pre-measured state of the measured system. 
The state (description) of the system, in a particular run of the experiment, is assumed to contain ``hidden variables'' that possess all the data necessary to dictate the values of all observables measured or could be measured in that run~\cite{lukono}.\\

%

\noindent \textbf{The locality assumption} states that the measurement outcomes of an observable in one system is independent of measurement settings in another system at a distant location~\cite{char}.\\

It is important to realize that quantum mechanics remains agnostic with respect to both the premises (of reality and locality). An observable that has not yet been measured has no outcome, except in the trivial case when the input belongs to one of the eigenspaces of the observable~\cite{durer-shuktara}. 
And, for the same reason, what would have been the value of \(A\) measured on a certain system, if the measurement on another (possibly distant) system is \(C\) instead of \(B\), in the current run of the experiment in which actually \(A\) and \(B\) were measured, has no answer within quantum mechanics.

For definiteness, we consider two observers having one subsystem each of a two-party system. The observers are traditionally called Alice and Bob~\cite{jome-doi}. We also suppose that Alice is measuring either the observable \(A\) or \(B\), while Bob has the choice to measure either \(B\) or \(C\).
We furthermore assume that all the three observables are dichotomic ones (i.e., has two outcomes each), and that the corresponding outcomes  
are represented by the real numbers, \(+1\) and \(-1\). 

Such observables do exist in the quantum realm, and in particular, spin measurements - ignoring the \(\frac{1}{2}\hbar\) factors - on quantum spin-1/2 systems are of this type.

The 1964 paper of Bell uses an additional assumption, just for the sake of definiteness. As we will see later, this is not required to obtain the Bell theorem. The assumption is that when the same observable is measured at both locations, the outcomes are anticorrelated. E.g., if \(A\) is measured by both Alice and Bob, then the outcomes are either \(+1\) at Alice and \(-1\) at Bob, or the other way round. 

Again, quantum theory allows the existence of states for which such anticorrelations occur, and the two-spin-1/2 singlet state is an example.

Using the reality assumption, we suppose that in a particular run of the experiment on the two-party state in possession of Alice and Bob, in which Alice measures \(A\) and Bob measures \(B\), the outcomes are denoted by \(a_A(\lambda)\) and \(b_B(\lambda)\). Here \(\lambda\) denotes the totality of all hidden variables (or just variables) needed for knowing the outcomes of all measurements that can possibly be performed on the two parts of the two-party system. 
\emph{In that same run,} if Bob had measured \(C\) instead of \(B\), then he would have obtained \(c_B(\lambda)\), and Alice would have still gotten \(a_A(\lambda)\).
Crucially, the locality assumption is already implicit here, as Alice's outcome is independent of whether Bob measures \(B\) or \(C\). The situation is similar if Alice and Bob measures \(B\) and \(C\) respectively. 

Let us assume that the hidden variable \(\lambda\) has a probability distribution \(\rho(\lambda)\) for the state preparation that we are considering in the experiment at hand. The correlator, \(\langle AB \rangle\), corresponding to measurements of \(A\) and \(B\) measured respectively at Alice and Bob is given by~\cite{Gadsisar}
\begin{equation}
\langle AB\rangle = \int d\lambda \rho(\lambda) a_A(\lambda) b_B(\lambda).
\end{equation}
The correlators \(\langle AC\rangle\) and \(\langle BC\rangle\) can be similarly defined. 

We therefore have
\begin{equation}
\langle AB\rangle - \langle AC\rangle
=\int d\lambda \rho(\lambda) a_A(\lambda)\left(b_B(\lambda) - c_B(\lambda)\right),
\end{equation}
where the assumption of reality is explicit, while the assumption of locality is used when we denote the outcome of a measurement of \(A\) by Alice with \(a_A(\lambda)\), irrespective of whether Bob measures \(B\) or \(C\). 
Therefore,
\begin{equation}
\langle AB\rangle - \langle AC\rangle
=\int d\lambda \rho(\lambda) a_A(\lambda)b_B(\lambda)\left(1 - b_B(\lambda)c_B(\lambda)\right),
\end{equation}
since \(b_B(\lambda)^2 =1\),
so that 
\begin{equation}
\left|\langle AB\rangle - \langle AC\rangle\right|
\leq \int d\lambda \rho(\lambda) \left(1 - b_B(\lambda)c_B(\lambda)\right),
\end{equation}
since \(|a_A(\lambda)b_B(\lambda)| =1\) and \(1 - b_B(\lambda)c_B(\lambda)\) is non-negative. Invoking the anticorrelatedness of the system state, we have 
\begin{equation}
\left|\langle AB\rangle - \langle AC\rangle\right|
\leq \int d\lambda \rho(\lambda) \left(1 + b_A(\lambda)c_B(\lambda)\right),
\end{equation}
i.e., 
\begin{equation}
1+\langle BC \rangle \geq \left|\langle AB\rangle - \langle AC\rangle\right|.
\end{equation}
This is the \textbf{Bell's inequality}.

A crucial point to note is that  the inequality has been derived without any reference to quantum theory, and is true for any physical theory for which assumptions of locality and reality hold, plus the additional assumption of anticorrelated outcomes. 

However, we can try to see whether quantum mechanics satisfies this inequality for the singlet state, 
\begin{equation}
\left|\psi^-\right\rangle = \frac{1}{\sqrt{2}}\left( 
\left| \uparrow_z\right\rangle 
\left| \downarrow_z\right\rangle - 
\left| \downarrow_z\right\rangle 
\left| \uparrow_z\right\rangle
\right),
\end{equation}
for arbitrary choice of the observables \(A\), \(B\), and \(C\). Here, \(\left| \uparrow_z\right\rangle \) and
\(\left| \downarrow_z\right\rangle\) are spin-up and spin-down states in the \(z\) direction of a quantum spin-1/2 system. Identifying \(A\), \(B\), and \(C\) with measurements of spin components in the directions \(\hat{a}\), \(\hat{b}\), and \(\hat{c}\), with \(\hat{a}\), \(\hat{b}\),  \(\hat{c}\) being three-dimensional unit vectors, we have 
\(A = \vec{\sigma} \cdot \hat{a}\), etc., where \(\vec{\sigma}\) is the triad of the three Pauli matrices. 
Consequently, for the singlet state,
\begin{equation}
\langle AB \rangle = - \hat{a}\cdot \hat{b},
\end{equation}
etc., so that the Bell inequality, if it holds for the singlet state, will imply
\begin{equation}
1-\hat{b}\cdot \hat{c} \geq \left|\hat{a}\cdot(\hat{b}-\hat{c})\right|.
\end{equation}
Note that we have here explicitly used the Born rule of quantum mechanics.
Choosing \(\hat{a} = (0,0,1)\), \(\hat{b} = (\sin\theta,0,\cos\theta)\), and \(\hat{c} = (\sin\theta{'} \cos \phi, \sin\theta{'} \sin \phi, \cos \theta{'})\), where \(\theta, \theta{'} \in [0,\pi/2]\) and \(\phi \in [0,\pi)\), we have
\begin{equation}
1-\sin\theta \sin \theta{'} \cos \phi - \cos \theta \cos \theta{'} \geq \left| \cos \theta - \cos \theta{'}\right|.
\end{equation}
Arbitrarily setting \(\phi =0\), we get 
\begin{equation}
1- \cos(\theta - \theta{'}) \geq \left| \cos \theta - \cos \theta{'}\right|.
\end{equation}
As can be easily checked, e.g. by plotting the LHS minus RHS on the \((\theta, \theta{'})\)-plane, this inequality is not always satisfied. 

Therefore, there exists quantum states and quantum measurements for which the Bell inequality can be violated. In other words, there exists quantum states and measurements for which the conjuction of locality and reality are not satisfied. This is the statement of the celebrated \textbf{Bell theorem}.

\section{Some further aspects of the paper}


The article of Bell mentions in the beginning that the reality assumption
had been declared to be  inconsistent with quantum mechanics even without the locality assumption, but that the proof has been shown to be wrong. The mistake  was detected
by Grete Hermann~\cite{ardhek-akash,bisleshan} and later by John Bell himself. Indeed, hidden variable theories of quantum mechanics are known to exist (e.g.,~\cite{bolo-horibol}), which however must all violate the locality assumption, as the 1964 paper of Bell requires.

The paper also provides a hidden variable theory for a single quantum spin-1/2 system.

\section{The CHSH inequality}
\label{sec-pile-chomkano}

The assumption of anticorrelation of certain measurement outcomes for the system state is a stringent one in an actual experiment as well as in applications. This can be removed e.g. by considering four measurement settings, as in the paper of Clauser, Horne, Shimony, and Holt (CHSH)~\cite{Shervani}, instead of the three settings considered in the paper of Bell.

Suppose therefore that our two observers, Alice and Bob, are in possession of a two-party state, and that Alice is able to measure either of two observables \(A\) and \(A{'}\), while Bob is able to measure either the observable \(B\) or the observable \(B{'}\). As before, we assume that these are dichotomic observables and that the outcomes of each of them are \(\pm 1\). 

It is now possible to derive an inequality - the CHSH or the Bell-CHSH inequality - that must be satisfied by any physical theory of the experimental setup in the preceding paragraph, and it is 
\begin{equation}
\left| \langle AB \rangle + \langle AB{'} \rangle + \langle A{'}B \rangle - \langle A{'}B{'} \rangle\right| \leq 2.
\end{equation}
Let us repeat that the derivation of the CHSH inequality does not require that the outcomes of certain observables measured on the system state are anticorrelated. 

Again, there are quantum states and quantum measurements for which this inequality is violated. In case of  maximal violation that can be obtained by any quantum state for any measurement settings, the LHS of the CHSH inequality reaches  \(2\sqrt{2}\)~\cite{Dehlieez}, and is attained for certain measurements on the singlet state.



\section{In lieu of a conclusion}

In lieu of a conclusion, let us first mention that there are a large variety of generalizations of the Bell inequality, each bringing with it certain new aspects, like ease of experimental realization, a new conceptual point, a new application, etc~\cite{gojikoT}. 

Secondly, it turned out to be rather difficult to experimentally demonstrate a violation of a Bell inequality, as a typical experimental evidence is often countered by ``loopholes'' in the experimental strategy~\cite{wanDapaTang}. 

Finally, a large variety of research areas, including ones with tantalizing applications, have prospered from relations with 
the concepts related to Bell's inequalities and the Bell theorem. An important example is the concept of entanglement witnesses, which currently forms one of the most sought-after methods to detect entanglement - a type of quantum correlation that is necessary though often not sufficient to violate a Bell inequality - in real experiments~\cite{veg-biriyani}. Another application of the Bell theorem is the possibility of obtaining device-independent security in cryptography~\cite{khide}.
 


\begin{thebibliography}{99}
\bibitem{lukono} In recent literature, hidden variables have often been referred to as just ``variables''.
\bibitem{char} Notice that the locality assumption implies that the Bell theorem needs the total system to be of at least two subsystems, which implies in particular that the total system is at least four-dimensional.
\bibitem{durer-shuktara} For an observable \(A\), the eigenspace corresponding to an eigenvalue \(a\) of \(A\) is the linear span of the eigenvectors of \(A\) corresponding to \(a\). If \(a\) is a non-degenerate eigenvalue of \(A\), then the eigenspace of \(a\) is one-dimensional.
\bibitem{jome-doi} A friend once pointed out that they could also be called Akbar and Birbal. Some younger friends mentioned some more pairs of options, but let us stick to Alice and Bob.
\bibitem{Gadsisar} This is referred to by a notation akin to \(P(A,B)\) in Bell's paper.
\end{thebibliography}
\end{document}